\documentclass[runningheads]{llncs}
\usepackage{graphicx}
\usepackage[hyphenbreaks]{breakurl}
\usepackage[hyphens]{url}
\usepackage{moreverb,url}

\usepackage[colorlinks,bookmarksopen,bookmarksnumbered,citecolor=black,urlcolor=black]{hyperref}

\newcommand\BibTeX{{\rmfamily B\kern-.05em \textsc{i\kern-.025em b}\kern-.08em
		T\kern-.1667em\lower.7ex\hbox{E}\kern-.125emX}}

\usepackage{graphicx}

\usepackage{smartdiagram}
\smartdiagramset{set color list={
		gray!30,
		gray!30,
		gray!30,
		gray!30,
		gray!30,
		gray!30,
	},
	sequence item border color=black,
	sequence item font size=16,
}

\tikzset{planet/.append style={
		draw=blue,thick
	},
	satellite/.append style={
		draw=red
	},
}

\usepackage{float}

\usepackage{color}
\usepackage{colortbl}

\usepackage{amssymb}
\usepackage{comment}
\usepackage[multiple]{footmisc}
\usepackage{amsmath}

\usepackage{lipsum}
\usepackage{pfnote}
\usepackage{multirow}

\usepackage{pgfplots}
\usepackage{tikz}
\usetikzlibrary{arrows,decorations.pathmorphing,fit,positioning}
\usepackage{xcolor}
\usetikzlibrary{patterns}
\usepackage{amstext}

\usepackage{fixfoot} 
\DeclareFixedFootnote{\myfnone}{one}  

\usepackage{enumitem}

\usepackage{colortbl}

\usepackage{bbding} 

\definecolor{maroon}{cmyk}{0,0.87,0.68,0.32}

\usetikzlibrary{positioning,calc}

\usepackage{tikz}
\usetikzlibrary{trees}

\tikzset{level 1/.style={level distance=2cm, sibling distance=10cm}}
\tikzset{level 2/.style={level distance=2cm, sibling distance=3cm}}

\tikzset{bag/.style={text width=10em, text centered,yshift=-0.5cm}}

\usepackage{smartdiagram}
\usesmartdiagramlibrary{additions}

\usetikzlibrary{positioning,calc}

\def\addlegendimage{\csname pgfplots@addlegendimage\endcsname}

\usepackage[utf8x]{inputenc}
\usepackage{smartdiagram}
\usepackage[english]{babel}

\makeatletter
\pgfkeys{/smart diagram/.cd,%
	uniform connection color/.is choice,%
	uniform connection color/true/.code={%
		\tikzset{connection planet satellite/.append style={%
				draw=\sm@core@connectioncolor
			},%
		}%
	},%
	uniform connection color/false/.style={connection planet satellite},%
	uniform connection color/.default=false,%
}%
\makeatother

\begin{document}
\title{An Exploratory Study of (\#)Exercise in the Twittersphere }
%
%
\author{George Shaw\inst{1}\orcidID{0000-0001-5508-3139} \and
Amir Karami\inst{2}\orcidID{0000-0003-1936-7497} }
\authorrunning{G. Shaw and A. Karami}
%
\institute{University of North Carolina at Charlotte, Charlotte NC 28223, USA \\
	\email{gshaw11@uncc.edu}\\
	University of South Carolina, Columbia SC 29208, USA\\
	\email{karami@sc.edu}}
\maketitle              
\begin{abstract}
Social media analytics allows us to extract, analyze, and establish semantic from user-generated contents in social media platforms.  This study utilized a mixed method including a three-step process of data collection, topic modeling, and data annotation for recognizing exercise related patterns.  Based on the findings, 86\% of the detected topics were identified as meaningful topics after conducting the data annotation process.  The most discussed exercise-related topics were physical activity (18.7\%), lifestyle behaviors (6.6\%), and dieting (4\%).  The results from our experiment indicate that the exploratory data analysis is a practical approach to summarizing the various characteristics of text data for different health and medical applications.    

\end{abstract}
\section{Introduction}

Surveys have been the traditional method used to gain insight into the complex understanding of patient satisfaction, opinions regarding a specific event, or assisting with decisions that should be made \cite{karami2018mining,varanasi2015seeking}.  However, there has been an increased interest in social media to collect information and enhance the data that is collected through traditional surveys \cite{benton2016collective,kitzie2018life}.  From the health perspective, social media have been used to investigate the health-related issues such as influenza \cite{culotta2010towards}, patient-doctor communication for assessing the quality of care \cite{hawn2009take}, diet, diabetes, and obesity \cite{karami2018characterizing,shaw2017computational}, and LGBT health \cite{karami2018characterizing,webb2018characterizing} on Twitter. 

Harnessing the opinions of a demographic group - or the general public – by using text mining methods within social media can help with understanding the phenomena under investigation. This also include the discovery of hidden topics \cite{varanasi2015seeking,karami2018us,collins2018social,karami2019political,karami2018computational}.  The applicability of text mining for content analysis and latent insight transcends multiple domains \cite{karami2014exploiting,karami2014improving1,karami2015online1}.  More importantly, the aforementioned studies identify the utility of the information collected. Likewise, the decision for individuals to exercise has implications for businesses, political realms, and healthcare.  A weight-loss focused study sought to identify the temporal trends in weight loss-related posts \cite{turner2015tweet}. Their results showed that more posts regarding weight loss were happening during the holidays; which is a period of the calendar year where weight gain commonly occurs.  For fitness companies, this allows them to increase the specificity of their marketing strategies when targeting new clients for their fitness center.  This also helps with understanding the granularity of sentiments, that are quite a context and geographically complex \cite{pang2008opinion}.  
Individuals within the political sphere that are running on a platform of reducing childhood obesity and the development of recreational centers can benefit from networking modeling and message engagements tactics using Twitter \cite{harris2014communication}.  Also, geographic locations with few mentions of exercising related topics can assist healthcare organizations aimed at preventative healthcare.  One research identified that physical activity (exercise) references were less frequent in census tracts that have greater economic disadvantages and higher proportions of racial/ethnic minorities and youths \cite{nguyen2016leveraging}.

While seemingly insignificant as a topic itself exercise through computational approaches has clear implications for how political campaigns are conducted, fitness facilities decision to build in a particular area, and strategies to increase physical activity through exergaming.  A significant portion of the current research has examined exercising through topic modeling approaches within a specific context; however, this research study seeks to explore the overall undertone of exercise within the Twitter environment.  

This paper sought to look at the broader context of the query term within Twitter and the utility of the information provided through a mixed-method approach using both computational and qualitative methods.  This research also addresses the practical implications for healthcare facilities and health professionals.  The following section discusses the methodology used in this exploratory study.

\section{Methodology and Results}

\subsection{Data Collection}
A real-time data collection method using the real-time Twitter Application Programming Interface (API) was used for this study .  From June 2016 to August 2016, a total of 3.7 million tweets were collected in English language using ``exercise" and ``\#exercise" queries.  To focus on just personal experiences, we removed the retweets and the tweets containing a URL and found 522,307 tweets.  Also, additional stop words and single characters such as ``and", ``or", and ``the" were removed.

\subsection{Topic Discovery and Analysis}

Topic models semantically cluster related words \cite{karami2018fuzzy,karami2017taming}. While there are various models for topic discovery, different studies showed the utility of the most cited topic model, Latent Dirichlet Allocation model (LDA) \cite{miller2017people,karami2014fftm,karami2015fuzzy,karami2015afuzzy}. LDA operates under the assumption that the words within a document reflect latent topics \cite{karami2018characterizing2,karami2015flatm}.  To identify the optimum number of topics, the log-likelihood estimation method was used.  Based on the log-likelihood, the optimum number of topics was 150. 

Using the qualitative coding method,  topics were considered relevant (identified) if they referenced exercising activities such as running, rock climbing, swimming, and Zumba.  Topics were also kept if they were deemed productive and counter-productive to engaging in exercising because participating in exercising requires a change of lifestyle behaviors \cite{wing2001behavioral}.

\vspace{-5mm}

\begin{table}[ht!]
	\scriptsize
	\centering
	\caption{Examples of the Identified and Unidentified Topics}
	\label{fig:topics}
	\begin{tabular}{|p{2cm}|p{1.7cm}|p{1.7cm}|p{1.9cm}|p{1.9cm}|p{2cm}|}
		\hline
	\rowcolor{maroon!10}	\textbf{Unidentifiable Topics}       & \textbf{Physical Activity}           & \textbf{Lifestyle Behaviors}  & \textbf{Adjectives}   & \textbf{Dieting}         & \textbf{Politics}           \\ \hline
	tayo  & exercise  & exercise    & true  & weight     & trump                         \\
		din                        & fit                         & health              & fat               & lose              & vote  \\
		pa                         & motivation                  & good                & lazy              & diet              & realdonaldtrump         \\
		ang                        & gym                         & sleep               & big               & loss              & hillary                            \\
		ng                         & fitness                     & lifestyle           & bigger            & eat               &    hillaryclinton                           \\
		umaga                      & active                      & habits              & flat              & gaining           &      campaign                           \\
		para                       & workout                     & nutrition           &                   & pounds            &               gop                \\ \hline
		
		Endorphins  & fitness                     & diet                & good              & calories   & restraint                     \\
		understand                      & workout             & plan                & bad               & minutes  & pr                            \\
		kill       & exercise                          & routine             & thing             & myfitnesspal               & forces                        \\
		shoot   & gym  & results             & feel              & week              & india                         \\
		makes  &    cardio                         & change              & pretty            & diet                 & power                         \\
		husbands  &      motivation                          & meal                & shape             &     plan              & govt                          \\
		groups &          training                   & schedule            & terrible          &        cals           & security                      \\ \hline
		
		pretty                     & exercises           & body                & fat               & diet              & democracy                     \\
		thing                      & exercise                       & mind                & skinny            & eating            & eu                            \\
		job                        & cardio                          & exercise            & hate              & weight            & brexit                        \\
		lol                        & abs             & reading             & lazy              & pounds            & uk                            \\
		oneself                    & training                      & soul                & body (bodily)     & lose              & vote                          \\
		yeah                       &  workout                            & healthy             & telling           & gained            & people              \\
		tho                        &         fitness                    & clear               & healthy           & kg                &        referendum \\ \hline                      
	\end{tabular}
\end{table}

Overall, of the detected 150 topics, 86\% (129) were identified as meaningful topics during the data annotation process (Table \ref{fig:topics}). Topics that constructed a significant portion of the dataset include physical activity (18.7\%), lifestyle behaviors (6.6\%), and dieting (4\%). These three health-related topics are of¬ten associated with exercising.  Dieting is an undervalued attribute for weight loss and optimizing physical activity performance. Surprisingly, diabetes was annotated as a topic once.  An intriguing topic was physiology, which is not an often-studied topic in social analytics related research.  Specifically, this is an excellent opportunity to explore gerontology associated issues as older adults increasingly use social media.  Older adults ``have used it to start discussion groups for life and health issues pertinent to them"\footnote{\url{https://www.huffingtonpost.com/anita-kamiel-rn-mps/older-people-social-media\_b\_9191178.html}}.   

\begin{figure}[ht!]
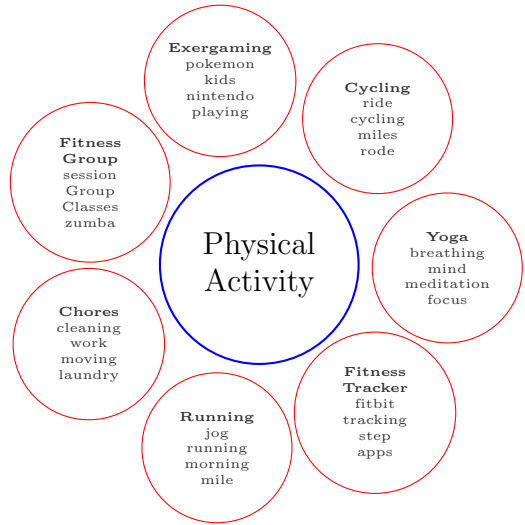

	\tiny
	\begin{center}
		\smartdiagram[constellation diagram]{Physical Activity,
			\textbf{Cycling} \\
			ride \\
			cycling \\
			miles \\
			rode
			, \textbf{Exergaming} \\
			pokemon \\
			kids \\
			nintendo \\
			playing
			, \textbf{Fitness Group} \\
			session \\
			Group \\
			Classes \\
			zumba
			,   \textbf{Chores} \\
			cleaning \\
			work \\
			moving \\
			laundry, 
			\textbf{Running} \\
			jog \\
			running \\
			morning \\
			mile,
			\textbf{Fitness Tracker} \\
			fitbit \\
			tracking \\
			step \\
			apps,
			\textbf{Yoga} \\
			breathing \\
			mind \\
			meditation \\
			focus
		}
		
	\end{center}
	\caption{Example of Sub-Level Labels for Physical Activity and Charactering Words}
	\label{fig:activity} 
\end{figure}

Examining the frequency of top exercising activities in the topics shows that walk (15), gym (13), and were the top three mentioned exercising activities. Semantically, topics of running and ran were included with the overall exercise activity of run. Surprisingly, exergaming (5) was listed fourth on the top exercising activities.  While the game achieved much popularity during the summer of 2016 \cite{leblanc2017pokemon}, these results may also contribute to the potential that augmented reality has with weight-loss interventions in public health.  The remaining top three exercise activities include Biking (5), Cardio (3), and Swimming (2). The results from this experiment identify the complexity and various forms of engaging in exercising To increase the specificity of topics clusters for physical activity, multiple sublevels of physical activity were identified. Among the sub-levels (Fig. \ref{fig:activity}), exergaming, fitness tracker and chores (domestic chores) are the most noticeable ones. 

\vspace{-5mm} 

\smartdiagramset{
	planet text width=2.5cm,
	bubble text opacity = 1,
	uniform connection color =true,
	connection color = none,	
	planet size=2cm,
	planet color = none,
	distance planet-text=0.1,
	distance planet-satellite=2.5cm,
	satellite fill opacity=0,
}

When examining the characteristics of each sub-level topic, each topic fits with the characteristics. For example, cleaning, work, moving, laundry, and kitchen, describes the activities that are typically conducted by an individual performing domestic chores.

\section{Conclusion}
The results from our experiment provided evidence that social media analytics regarding exercising is a complex subject among Twitter users.  This research provided an in-depth analysis of the data collected that is not often achieved with traditional qualitative methods.  Therefore, it is important to identify the researchers' limited expertise as health experts and future studies should include medical or health experts to annotate the topics.  We also think future research studies should compare multiple social media platforms, such as Facebook and Reddit, to examine the topics across platforms and how they are used by health organizations to disseminate health information regarding exercising.

%
%
%
\bibliographystyle{splncs04}
%
\bibliography{refrence}
\end{document}